\input{aipcheck}

\documentclass[
    ,final            
  ]
  {aipproc}

\layoutstyle{8x11double}

\begin{document}

\title{Coexistence of triplet superconductivity and itinerant ferromagnetism}


\author{V.P.Mineev}{address={Commissariat a l'Energie Atomique, INAC/SPSMS, 38054 Grenoble, France}
}

\begin{abstract}
The triplet superconductivity in  $UGe_2$ and $URhGe$ coexists with itinerant ferromagnetism
such that in the pressure-temperature phase diagram the whole region occupied by the superconducting state is situated inside a more vast ferromagnetic region.
In the same family metal $UCoGe$ the pressure dependent 
critical lines $T_{Curie}(P)$ and  $T_{sc}(P)$ of the ferromagnet and the superconducting phase transitions intersect each other. The two-band multidomain  superconducting ferromagnet state arises at temperatures below both of these lines.  Here I describe 
the symmetry and the order parameters of the 
paramagnet as well of the multidomain ferromagnet superconducting states.
The Josephson coupling between two adjacent  ferromagnet superconducting domains is discussed.

\end{abstract}

\maketitle


\section{Introduction}

A phase transition of the second order breaks some symmetry such that below the critical temperature 
the ordered phase of lower symmetry in comparison with the initial state is formed. As it was first  
pointed out by L.D.Landau \cite{Lan}
an intersection of critical lines on the phase diagram leads to formation of an ordered  phase with symmetry lower than the symmetries of both  initial ordered states existing below of each critical lines separately. Here we study the symmetries of ordered phases arising at
intersection of critical lines of ferromagnet and superconducting phase transitions.

 The co-existence of superconductivity and ferromagnetism in several uranium compounds
 $UGe_2$, \cite{Saxena} $URhGe$, ~\cite{Aoki} and the recently revealed $UCoGe$.~\cite{Huy07} is found to arise as a co-operative phenomena rather than as the overlap of two-mutually competing orders. In all these compounds the substantial reduction of the ordered moment as compared with the Curie-Weiss moment provides clear evidence of $5f$ itineracy.
 In the first two compounds the Curie temperatures $T_{Curie}$ is more than the order of magnitude higher than their critical temperatures for superconductivity. In $UCoGe$ the ratio $T_{Curie}/T_{sc}$ at ambient pressure  is about four.
 The large exchange field and also high   upper critical field at low temperatures  strongly exceeding the 
 paramagnetic limiting field ~\cite{Huxley01,Hardy051,Huy08} 
 indicate that here we deal with Cooper pairing in the triplet state.
\begin{figure}
\includegraphics[height=.5\textheight]{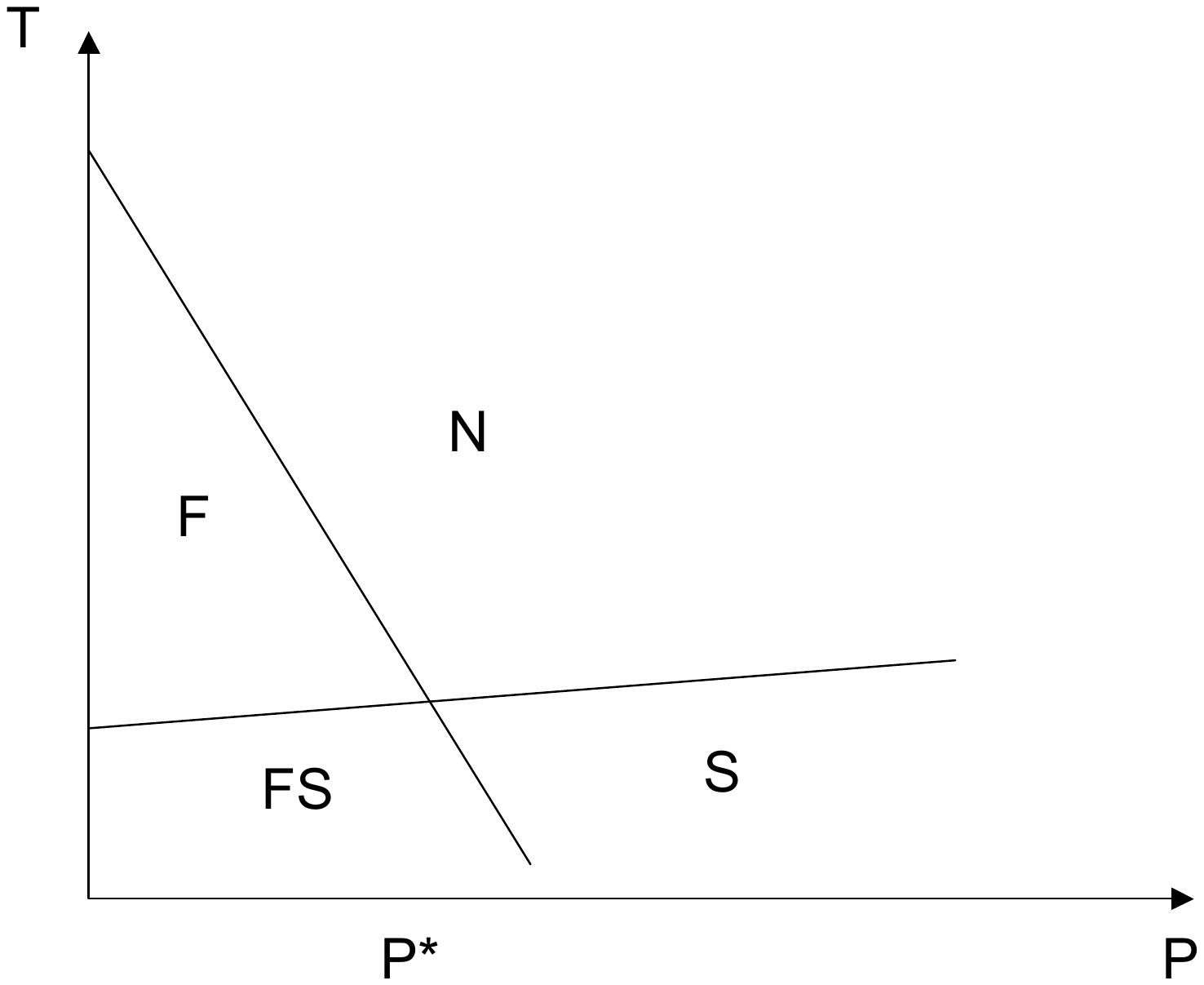}
 \caption{The schematic pressure-temperature phase diagram of superconducting $UCoGe$.
  Here, $N$ is the normal paramagnet phase, $F$ is the ferromagnet phase, $S$ is the paramagnet superconducting phase, $FS$ is the multi-domain ferromagnet superconducting phase. All the lines are the lines of the second-order phase transitions.}
\end{figure}
 
The singlet superconductivity  coexists
with ferromagnetism in a form  known
as the Anderson-Suhl or cryptoferromagnetic superconducting state (for review see \cite{Buzdin}) characterized by the formation of a transverse domain-like
magnetic  structure. The structure period or  domain size is larger than interatomic distance and smaller than the superconducting coherence length that weakens the depairing effect of the exchange field.
The latter is irrelevant in the case of triplet superconductivity. Hence, there is no reason for the formation of a cryptomagnetic state.  Indeed, no traces of space modulation of magnetic moments directions
on the scale smaller than the coherence length has been revealed.\cite{Aoki,Aso5,Kotegawa,Ohta} 
On the other hand the neutron depolarization measurements on $UGe_2$ down to 4.2 K  (that is in the ferromagnet but not superconducting region) establish, that the magnetic moment strictly aligned along a-axis, with a typical domain size in the bc-plane of the order $4.4\times10^{-4}$  cm \cite{Sakarya}
that is about two orders of magnitude larger than the largest superconducting coherence length in b-direction $\xi_b\approx 7\times10^{-6}$ cm.

 Arising at temperatures far below the corresponding Curie temperature the superconductivity in $UGe_2$ and $URhGe$   coexists with ferromagnetism in some pressure interval such that in the $(P,T)$ phase diagram the whole   region occupied by the superconducting state is situated inside a more vast ferromagnetic region.\cite{Huxley03,Hardy052}  In $URhGe$  the Curie temperature increases up to the highest pressure achieved $(130~ kbar)$. The superconducting critical temperature decreases slowly up to 20 kbar. It is more peculiar the behavior of $UGe_2$, where at low temperatures the ferromagnetism and the superconductivity abruptly ( by means the first-order-type transition) disappears  at the same critical pressure $P_c\approx 15~ kbar$. 
 
The observation that the superconductivity in $UGe_2$ is confined to the ferromagnet state can be trivially explained by an assumption that the ferromagnetism in this compound is formed by f-electrons with half-metallic bands filling. Namely, the band with the spin-down electrons is completely filled, whereas the band with spin-up electrons filled up to the Fermi level. The triplet spin-up superconducting state formed in this band persists so long the Fermi level intersects this band. The pressure induced the Fermi level lifting above the band upper boundary kills both the itinerant ferromagnetism and the superconductivity.  

 The particular one-band superconducting state was chosen
\cite{Hardy051} for successful explanation of temperature dependences of the upper critical field in $URhGe$ in different crystallographic directions. This state is also appropriate for the description  \cite{Mineev2006} of the transition driven by the change of orientation of the ordered magnetic moment in this compound by the application of magnetic field in perpendicular direction accompanying by the arising a reentrant superconducting state \cite{Levy}.
The one band superconductivity, of course, does not exclude the existence of the other conducting but not superconducting bands, or, more exactly, the bands with negligibly small superconducting gaps. The latter is in correspondence with reduced specific heat jump in comparison with BCS value, and the finite residual zero temperature ratio 
$(C(T)/T)_{T\to0}$ comparable with its magnitude in the normal state found in all uranium superconducting ferromagnets.

  The phase diagram of  the new 
ferromagnetic superconductor $UCoGe$  is qualitatively different (see Figure 1).\cite{Hassinger}  At ambient pressure, the ferromagnetism ($T_{Curie}\approx 3 K$) coexists with superconducting state  ($T_{sc}\approx 0.7 K$). Then at applied pressure,  the Curie temperature decreases 
such that no indication of ferromagnetic order is observed above $P^*\approx 10~kbar$. The resistive superconducting transition is, however, quite stable with changes in temperature and persists up to the highest measured pressure of about $24~ kbar$.  
Thus, the pressure dependent transition lines $T_{Curie}(P)$ and $T_{sc}(P)$ apparently intersect each other and 
the superconductivity exists both in the paramagnet and in the ferromagnet state. 

The ferromagnet superconducting state in an orthorhombic  metal is similar to the superfluid $^3He-A$ in an external magnetic field known as $A_2$ state.
The superfluid 
$^3He-A$ is the spin nonpolarized  state   formed  by the spin-up and the spin-down Cooper pairs in equal amounts. There is also the spin-polarized  $A_1$ state  
where the pairing only spin-up particles occurs.\cite{Leggett} The  $^3He-A_1$ arises from the normal Fermi liquid in an external magnetic field. Then, at lower temperature, the liquid passes to the $A_2$ 
state where the paired spin-up and spin-down states are almost equally populated.
The presence of  spin-orbital coupling admixes some amount of the spin-down Cooper pairs   to pure $A_1$ state \cite{Kojima}, such that the  $A_1$ and $A_2$ states are in fact qualitatively indistinguishable. The phase transition between these two states is a crossover, looking as a phase transition due to the smallness of the spin-orbital coupling in superfluid $^3He$.
The nonunitary  two-band superconducting state arising in the ferromagnet state of $UCoGe$ 
can be considered as an analog of superfluid $A_2$ phase arising from the  normal liquid
$^3He$  under magnetic field.

The increasing pressure causes the decrease of the exchange field that suppresses
the spin-up and spin-down band difference. 
The restoration of the time reversal symmetry occurs at recreation of spin-up and spin-down band degeneracy by the phase transition from the ferromagnet {\it axiplanar} superconducting state 
to the paramagnet superconducting state similar to the {\it planar} state of the superfluid $^3He$
(for the superfluid $^3He$  phase definitions see for instance \cite{MinSam}). 
Thus,   the ferromagnet superconducting state is separated from the normal state by the more symmetric, paramagnet planar-like state. We see, that the $(P,T)$ phase diagram in $UCoGe$ is quite naturally explained in terms of two band superconducting state in this material.
The observation of the upward curvature in the temperature dependence of the upper critical field 
in $UCoGe$ ~\cite{Huy08} 
adds the additional argument in support of this point.

The symmetries and  the 
order parameters of unconventional superconducting states
arising from the normal state with a ferromagnetic order in orthorhombic crystals with strong spin-orbital coupling have been found in the paper. \cite{Mineev} Then it was pointed out  that superconducting states in triplet ferromagnet superconductors represent a special type of two band superconducting states. \cite{Cham,Min04}.  There were obtained several results based on phenomenlogical (Ginzburg-Landau) and microscopic descriptions of two-band superconductivity.
It was proved, however, that  the superconducting ferromagnet classes  
pointed there have been found improperly.
Although, it leaves untouched the main results of  \cite{Mineev,Cham,Min04}, 
the based on these papers description  of possible $(P,T)$ phase diagrams for two band superconducting states in an orthorhombic itinerant ferromagnet  is incorrect.\cite{Min08} To make it correctly we  return  to the definition of the superconducting ferromagnet classes. It will be proven that $T_{Curie}(P)$ and
$T_{sc}(P)$ can intersect each other as  the critical lines of the phase transition of the second order.
The symmetry and the order parameters of the multidomain ferromagnet 
as well of the paramagnet superconducting states are established.
The Josephson coupling between neighboring superconducting domains is also discussed.

  \section{Two-band superconducting ferromagnet phase diagram}

All uranium ferromagnetic superconductors are  orthorhombic metals.
The symmetry of its  normal paramagnetic state is determined by the elements of the group
\begin{equation}
G_N=D_2\times U(1)\times R,
\label{e-1}
\end{equation}
where $D_{2}=(E, C_{2}^{z},
C_{2}^{x}, C_{2}^{y})$ is the point symmetry group including the operations    $C_{2}^{x}, C_{2}^{y},C_{2}^{z}$ of rotation on the angle $\pi$ about the $x,y,z$- axes correspondingly,   $U(1)$ is the group of gauge transformations, and 
$R$ is the time reversal operation. 

If the pressure dependent transition lines $T_{Curie}(P)$ and $T_{sc}(P)$ intersect each other at $P=P^*$ then the region of the coexistence of superconductivity and ferromagnetism is separated from the normal state by the region of ferromagnet normal state at $P<P^*$ and by the region of the superconducting state at $P>P^*$ (see Figure 1).

In the transition from the normal paramagnet state to the normal ferromagnet state the magnetic moment
directed along one crystallographic axis appears. We chose this direction as the $\hat z$ axis.
Hence, in the ferromagnet state
the  symmetry reduces to the 
\begin{equation}
G_F=D_2(C_2^z)\times U(1),
\label{e0}
\end{equation}
where
\begin{equation} 
D_2(C_2^z)=(E,C_2^z,RC_2^x,RC_2^y)
\label{e00}
\end{equation}
 is the so called {\it magnetic class} \cite{Landau} or the point symmetry group of the ferromagnet. The rotations on the angle $\pi$ about the $x$- and $y$- directions are accompanied by the time inversion $R$ that changes  the direction of magnetization 
to the opposite one.

In the transition from the normal paramagnet state to the superconducting paramagnet state the gauge symmetry is broken, such that the symmetry of this, so called {\it conventional} superconducting state is 
\begin{equation}
G_S=(E,C_2^z, C_2^x, C_2^y)\times R.
\label{es}
\end{equation} 
There is another possibility related to the formation of {\it nonconventional} superconducting state
where, in the addition to the gauge symmetry,  the point symmetry is also broken. We shall not discuss it here.

Now, we shall consequently describe the phase transitions from the normal ferromagnet state (F)  to the superconducting ferromagnet state (FS) taking place at $P<P^*$ and from the paramagnet superconducting state (S) to the superconducting ferromagnet state (FS) at $P>P^*$.

\subsubsection{F to FS phase transition}

As it was remarked in \cite{Cham}  superconducting state in an itinerant ferromagnet represents the special type of two band superconducting state  
 consisting of pairing states formed by spin-up electrons from one band and by spin-down electrons from another band. 
Hence, a superconducting state 
characterizes by two component order parameter  
\begin{equation}
{\bf d}_1({\bf k})=\Delta_{\uparrow}({\bf k})(\hat{x}+i\hat{y}),~~~
{\bf d}_2({\bf k})=
\Delta_{\downarrow}({\bf k})(\hat{x}-i\hat{y}).
\label{e1}
\end{equation}
Here, $\hat{x}$ and $\hat{y}$  are the
unit vectors of the spin 
coordinate system pinned to the crystal axes.  

The  unconventional superconducting states
arising from the normal state with a ferromagnetic order in orthorhombic crystals with strong spin-orbital coupling belong to the two different corepresentations $A$ and $B$.\cite{Mineev}  All the states relating to the given corepresentation obey the same critical temperature. 
The order parameter amplitudes  for $A$ and $B$ states correspondingly are given by
\begin{eqnarray} 
\Delta_{\uparrow}^A({\bf k})=\eta_1(k_xu_1+ik_yu_2),\nonumber\\
 \Delta_{\downarrow}^A({\bf k})=
\eta_2(k_xu_3+ik_yu_4),
\label{e2}
\end{eqnarray}
\begin{eqnarray} 
\Delta_{\uparrow}^B({\bf k})=\eta_1(k_zv_1+ik_xk_yk_zv_2),\nonumber\\
 \Delta_{\downarrow}^B({\bf k})=
\eta_2(k_zv_3+ik_xk_yk_zv_4).
\label{e2'}
\end{eqnarray}
They are odd functions of the momentum directions of pairing particles on the Fermi
surface. The functions
$u_i=u_i(k_x^2, k_y^2, k_z^2)$ and $v_i=v_i(k_x^2, k_y^2, k_z^2)$ are invariant in respect of all transformations of orthorhombic group. For the brevity, in that follows,  we shall discuss only the $A$ state. This state  is  related to the family of nonunitary axiplanar states.

The complex order parameter amplitudes $\eta_1=|\eta_1|e^{i\varphi_1}$
and $\eta_2=|\eta_2|e^{i\varphi_2}$ are not completely independent. The  relative phase difference $\varphi_1-\varphi_2$ is chosen such that the  quadratic in the order parameter part of the Ginzburg-Landau free energy density should be minimal.  
In an ordinary two-band superconductor it is 
\begin{equation} 
F=\alpha_1|\eta_1|^2+\alpha_2|\eta_2|^2+\gamma(\eta_1^*\eta_2+\eta_1\eta_2^*),
\label{44}
\end{equation}
and  $\varphi_1-\varphi_2=\pi$ for $\gamma>0$ and $\varphi_1-\varphi_2=0$ for $\gamma<0$.
In the case of ferromagnetic normal state the time reversal symmetry is broken and the quadratic in the order parameter components free energy density has the form 
\begin{eqnarray} 
F=\alpha_1|\eta_1|^2+\alpha_2|\eta_2|^2+\gamma(\eta_1^*\eta_2+\eta_1\eta_2^*)
\nonumber\\+
i\delta(\eta_1^*\eta_2-\eta_1\eta_2^*).
\label{45}
\end{eqnarray}
Here, all the coefficients are the functions of the exchange field $h$. 
The last term breaks the time reversal symmetry.     
In the absence of exchange field $\delta=0$. 
Minimization of  free energy \eqref{45} fixes the order parameter components phase difference $\tan(\varphi_1-\varphi_2)=\delta/\gamma$.  After substitution of this value back to \eqref{45} we come to the expression
\begin{equation} 
F=\alpha_1|\eta_1|^2+\alpha_2|\eta_2|^2
-\sqrt{\gamma^2+\delta^2}( \eta_1^*\eta_2+\eta_1\eta_2^*).
\label{freeen}
\end{equation}
Here $\alpha_i=\alpha_{i0}(T-T_{ci})$,  $i=1,2$ are the band indices, $T_{ci}$ are the critical temperatures in each band in the absence of band mixing.
Unlike eqn. \eqref{45} the complex amplitudes  $\eta_1=|\eta_1|e^{i\theta}$, $\eta_2=|\eta_2|e^{i\theta}$ in the eqn. \eqref{freeen} have common phase factors with $\theta=(\varphi_1+\varphi_2)/2$.
This form of free energy valid near the phase transition from the ferromagnet state to the ferromagnet superconducting state has been used in the papers. \cite{Cham,Min04} 
The common for the each band superconductivity critical temperature is given by
\begin{equation}
T_{sc}=\frac{T_{c1}+T_{c2}}{2}+\sqrt{\left (\frac{T_{c1}-T_{c2}}{2}\right )^2+\frac{\gamma^2+\delta^2}{\alpha_{10}\alpha_{20}}}
\end{equation}

In the superconducting $A$-state the gauge symmetry is broken. Acting on the order parameters (5), (6) 
by the elements $g$ of $D_2(C_2^z)=(E,C_2^z,RC_2^x,RC_2^y)$ group we obtain the following  
coefficients of transformation, or matrices of corepresentation 
\begin{equation}
\Gamma_1=(1,1, e^{-2i\varphi_1}, e^{-2i\varphi_1}),~~~
\Gamma_2=(1,1, e^{-2i\varphi_2}, e^{-2i\varphi_2}),
\label{Gamma}
\end{equation}
correspondingly.  Corepresentations $\Gamma_1$ and $\Gamma_2$ are equivalent  or they  are transformed each other by an unitary matrix $U$ as 
$\Gamma_1(g) =U^{-1}\Gamma_2(g)U$ if the element $g$ does not include the time inversion, and as 
$\Gamma_1(g) =U^{-1}\Gamma_2(g)U^*$ if the element $g$ includes the time inversion. It is easy to check that here the matrix of transformation is $U=e^{i(\varphi_2-\varphi_1)}$. 

The order parameter component ${\bf d}_1({\bf k})$ relating to  the spin-up band  is
invariant in respect to the following group of transformations 
\begin{equation}
G_{FS}=(E,C_2^z, RC_2^x, RC_2^y)=D_2(C_2^z).
\label{46}
\end{equation}
Action of the 
time reversal operation $R$ on    superconducting order parameter implies also the multiplication of it by the square of its phase factor: $R\to e^{2i\varphi_1}R$.
The second component ${\bf d}_2({\bf k})$ possess the same symmetry. So,  the group of symmetry of 
superconducting ferromagnet state $A$ called also by  {\it the superconductiing magnetic class} is $D_2(C_2^z)$. This group is the subgroup of the group of symmetry of the ferromagnet state \eqref{e0}.

\subsubsection{Superconducting ferromagnet domains}

The Cooper pairing changes the magnitude of spontaneous magnetization in respect to its value in normal ferromagnet state. Namely, the superconducting spin magnetic moment density is
\begin{equation}
{\bf M}_s=\mu_B\left [N'_{0\uparrow} \langle|\Delta_{\uparrow}¥({\bf k})|^2\rangle-N'_{0\downarrow}\langle|\Delta_{\downarrow}¥({\bf k})|^2\rangle \right].
\end{equation} 
Here, in the first term, $N'_{0\uparrow}$ is the derivative of the density of states at the Fermi surface of the spin-up band ,
and the angular brackets means the averaging over it. The second term presents the corresponding input of the spin-down band. One can write also the orbital magnetic moment density. \cite{MinSam}

Along with the introduced state $A$, there is its time reversed state $A^*$
characterized by the complex conjugate order parameter components 
\begin{eqnarray}
{\bf d}_1^*({\bf k})=\zeta_1(\hat{x}-i\hat{y})(k_xu_1-ik_yu_2),\nonumber\\
{\bf d}_2^*({\bf k})=\zeta_2(\hat{x} +i\hat{y})(k_xu_3-ik_yu_4).
\end{eqnarray}
The states 
$A$ and  $A^*$ occupy neighboring  domains with the opposite direction of magnetization.
The state $A^*$ order parameter amplitudes are $\zeta_1=|\zeta_1| e^{i\phi_1}$ and $\zeta_2=|\zeta_2| e^{i\phi_2}$. The phase difference is fixed by  $\tan(\phi_1-\phi_2)=\delta(-h)/\gamma$. 

The matrices of corepresentations for the state $A^*$ are obtained from \eqref{Gamma} by the substitution $\varphi_{1,2} \to \phi_{1,2}$. So, they transformed each other by means the matrices
$U_i=e^{i(\varphi_i-\phi_i)}$.   
It means, that the corepresentations for the state $A^*$ are  equivalent  to the corepresentations for the state $A$. Hence, the superconducting states in the neighboring domains obey the same critical temperature.

The symmetry of the time reversed states $A^*$ belong to the same 
 {\it superconducting ferromagnet class} $D_2(C_2^z)$ as the $A$-states.

\subsubsection{S to FS phase transition}

 In $P>P^*$ region  at temperature decrease $UCoGe$ pass to the nonmagnetic superconducting state.
 Let us assume the simplest and quite natural situation that it is the superconducting state  with the order parameter 
\begin{equation}
{\bf d}({\bf k})=2\eta(
k_xw_1\hat {x}+k_yw_2\hat y),
\label{d}
\end{equation}
transforming according to the unit representation of the normal state point symmetry group $D_2$. 
Here  $\eta=|\eta|e^{i\varphi}$ and the functions
$w_{1,2}=w_{1,2}(k_x^2, k_y^2, k_z^2)$ 
are invariant in respect of all transformations of orthorhombic group. This state reminds
planar phase of superfluid $^3He$. The paramagnet superconducting state is invariant in respect to the group \eqref{es} which can be rewritten as 
\begin{equation}
G_S=D_2(C_2^z)+R\times D_2(C_2^z).
\label{G_S}
\end{equation}

By further decrease the temperature
we approach to 
$T_{Curie}(P)$. At this temperature the exchange field appears, and the Kramers degeneracy between spin-up and spin-down electron states is lifted accompanied by arising of deviation from of the order parameter \eqref{d} 
\begin{eqnarray}
&{\bf d}({\bf k})=2\eta(
k_xw_1\hat {x}+k_yw_2\hat y)~~~~~~~~~~~~\nonumber\\ &=\eta(k_xw_1-ik_yw_2)(\hat x+i\hat y)+
\eta(k_xw_1+ik_yw_2)(\hat x-i\hat y)\nonumber
\\
&\to \tilde{\bf d}({\bf k})=\eta_1(k_xw_1-ik_yw_2)(\hat x+i\hat y)\nonumber\\ &+
\eta_2(k_xw_1+ik_yw_2)(\hat x-i\hat y)
\label{d'}
\end{eqnarray} 
The order parameter $\tilde{\bf d}({\bf k})$ transforms according to  corepresentation of the symmetry group
\eqref{G_S} of the paramagnet superconducting state. 

Along with increase of the band splitting the two component of the order parameter $\tilde{\bf d}({\bf k})$  are transformed to the order parameters of spin-up and spin-down bands given by eqns. \eqref{e1}, \eqref{e2}.
The ferromagnet superconducting state determined by eqn. \eqref{d'} as well  by  the eqns. \eqref{e1} and \eqref{e2} is invariant in respect to the group  
\begin{equation}
G_{FS}=D_2(C_2^z). 
\end{equation}
The latter
is the subgroup of the group of symmetry of ferromagnet state $G_F$ \eqref{e0} as well as of the symmetry group of paramagnet superconducting state $G_S$ \eqref{G_S}. So, the lines of the ferromagnet and the superconducting phase transitions can intersect each other as the critical lines of the phase transitions of the second order.



\section{Interdomain Josephson coupling}
Let us consider a
 flat domain wall   dividing magnetic moment-up and -down domains in single band ferromagnet.
 This case, the localized at $x=0$ domain wall contribution to the superconducting free energy density is given by  \cite{Sam} 
\begin{eqnarray} 
F_{DW}=\left [\gamma_1(|\eta|^2+|\zeta|^2)+\gamma_2(\eta^*\zeta+\eta\zeta^*)
\right.\nonumber\\+
\left.i\gamma_3(\eta^*\zeta-\eta\zeta^*)\right ]\delta(x).
\label{Fs}
\end{eqnarray}
Here $\eta=|\eta|e^{i\varphi}$ and $\zeta=|\zeta|e^{i\phi}$ are the superconducting order parameters in the left (magnetic moment-up) domain and in the right (magnetic moment-down) domain, correspondingly. The boundary conditions at $x=0$ are derived by the minimization of the sum of domain wall \eqref{Fs} and the gradient free energies.\cite{MinSam}
\begin{eqnarray} 
K\frac{\partial\zeta}{\partial x}=\gamma_1\zeta+(\gamma_2+i\gamma_3)\eta\nonumber\\
-K\frac{\partial\eta}{\partial x}=\gamma_1\eta+(\gamma_2-i\gamma_3)\zeta.
\label{bc}
\end{eqnarray}
Here, the rigidity coefficients $K\sim\hbar^2/m$.
The solutions of left and right domain  nonlinear Ginzburg-Landau equations supplemented by these boundary conditions determine the order parameter distribution of two domain superconducting structure. The solution of corresponding linear problem is physically relevant only in the case of stimulation of superconductivity by the domain wall when the localized near domain wall
superconducting state arises at temperatures higher than the temperature of superconducting phase transition in single domain geometry.

The situation for two band superconductivity is much more complicated.
This case the two-band 
domain wall free energy density is obtained by the addition to the eqn.\eqref{Fs}
the corresponding terms for the second band order parameters $\eta_2$ and $\zeta_2$ and also the interband terms symmetric in respect to the substitutions
$\eta_i$ by $\zeta_i$ and vice versa. 

Substituting   the boundary conditions \eqref{bc} in  the sum of the left domain and the right domain current through the domain wall (see for instance \cite{MinSam})  we obtain the density of the  interdomain Josephson current:
\begin{equation}
{\bf j}=\frac{8eK}{\hbar}|\zeta||\eta|\left [ \gamma_2\sin(\phi-\varphi)-\gamma_3\cos(\phi-\varphi)\right ]
\end{equation}
Thus, due to the time reversal breaking ($\gamma_3\ne0$) the expression for the Josephson current 
between the adjacent superconducting domains with spin-up and spin-down magnetization 
differs from the usual weak link Josephson current formula.  In the equilibrium, the phase difference between domains is fixed: $\tan(\phi-\varphi)=\gamma_3/\gamma_2$, and a spontaneous interdomain current is absent. 

In conclusion of this section  it is worth to be noted that the existence of the interdomain Josephson coupling bilinear in respect of $|\eta|$ and $|\zeta|$ is typical for the $A$ superconducting states. The order paramer for the $B$ states is vanishing in the equatorial plane $k_z=0$.  This case, there is  only the higher order  Josephson coupling between the domains divided by a flat domain wall parallel to the magnetization direction.

\section{Conclusion}

 The superconducting state in the itinerant ferromagnet uranium compound
 $UCoGe$
  manifests  the properties naturally explained in terms of two band superconductivity with triplet pairing. We discussed the symmetry and the order parameters of such a state.
It was proven that the pressure dependent critical lines of the ferromagnet $T_{Curie}(P)$ and the superconducting phase transition 
$T_{sc}(P)$ can intersect each other in the correspondence with the experimental observations.\cite{Hassinger} The Josephson coupling in between the adjacent superconducting ferromagnet domains is found to be different in comparison with the usual Josephson coupling between two superconductors.

\bibliographystyle{aipproc}   

\bibliography{sample}

\IfFileExists{\jobname.bbl}{}
 {\typeout{}
  \typeout{******************************************}
  \typeout{** Please run "bibtex \jobname" to optain}
  \typeout{** the bibliography and then re-run LaTeX}
  \typeout{** twice to fix the references!}
  \typeout{******************************************}
  \typeout{}
 }


\end{document}